# Relationship between Entropy and Diffusion: A statistical mechanical derivation of Rosenfeld expression for a rugged energy landscape

**Kazuhiko Seki**,[1] and **Biman Bagchi** [2 a)]

[1] *National Institute of Advanced Industrial Science and Technology (AIST), Tsukuba Higashi 1-1-1, 305-8565, Japan*

[2] *Solid State and Structural Chemistry Unit, Indian Institute of Science, Bangalore – 560012, India*

Diffusion – a measure of dynamics, and entropy – a measure of disorder in the system, are found to be intimately correlated in many systems, and the correlation is often strongly non-linear. We explore the origin of this complex dependence by studying diffusion of a point Brownian particle on a model potential energy surface characterized by ruggedness. If we assume that the ruggedness has a Gaussian distribution then for this model, one can obtain the excess entropy exactly for any dimension. By using the expression for the mean first passage time (MFPT), we present a statistical mechanical derivation of the well-known and well-tested scaling relation proposed by Rosenfeld between diffusion and excess entropy. In anticipation that Rosenfeld diffusion-entropy scaling (RDES) relation may continue to be valid in higher dimensions (where the mean first passage time approach is not available), we carry out an effective medium approximation (EMA) based analysis of the effective transition rate, and hence of the effective diffusion coefficient. We show that the EMA expression can be used to derive the RDES scaling relation for *any dimension higher than unity*. However, RDES is shown to break down in the presence of spatial correlation among the energy landscape values.

[a)] Corresponding author email: profbiman@gmail.com

# I. INTRODUCTION

Study of relationship between diffusion (D) of a tagged molecule and the entropy (S) of complex systems has a long history.[1-7] The first quantitative relation between the two was established by Adam and Gibbs in the following famous relation[4]

$$D(T) = D(T = T_0) \exp(C/TS_C) , \qquad (1)$$

where $S_c$ is the configuration entropy of the system, defined as

$$S_C(T) = S(T) - S_{Vib}(T) , \qquad (2)$$

where $S_{Vib}$ is the vibrational entropy of the system and $D(T_0)$ is the diffusion coefficient at a higher temperature $T_0$. The derivation of Adam-Gibbs contains certain critical concepts, like "cooperatively rearranging region" that has been used repeatedly by the scientist community and the theory seems to have survived the test of time, although a quantitative verification has proven to be elusive. But it is generally agreed upon that Eq. (1) correctly describes, qualitatively, the strong dependence of diffusion on entropy. An interesting relation of Eq. (1) has been given by Xia and Wolynes who invoked the idea of nucleation to explain slow relaxation in glassy liquids.[5] Another oft quoted expression is due to Rosenfeld who proposed the following relation between diffusion constant and entropy[6-8]

$$D = a \exp(bS_{ex}/k_B) , \qquad (3)$$

where $S_{ex}$ is the excess entropy of the system defined by subtracting the ideal gas entropy from the total entropy, and $a$ and $b$ are empirical fitting parameter. $k_B$ is Boltzmann's constant.

Although less well-known than the expression of Adam and Gibbs, Eq. (3) has been employed fairly often, mostly in the high temperature region. Recently a justification of Rosenfeld's expression has been discussed within a one dimensional model with rough energy landscape where it was shown that dependence on ruggedness of entropy and diffusion obeys Rosenfeld scaling with $a=1$ and $b=2$ which are also close to the values obtained by fitting to other systems. However, no quantitative derivation of Rosenfeld scaling is known, to the best of our knowledge. Given the different functional forms of Eqs. (1) and (3), it is surprising that both seem to describe entropy dependence of diffusion over a common range of temperature. It however seems safe to assume that Rosenfeld form holds better at high temperature while Adam-Gibbs is more suitable at lower temperature. It is thus an interesting question to inquire a crossover, if any, signifies a change in dynamics of the liquid.

In recent times, one often employs a description of collective dynamics of the system that is based on a rugged energy landscape.[9-20] In the present work we explore the relationship between entropy and diffusion in a system characterized by energy landscape that is rugged (or, rough), and establish certain definitive relations between the two.

## II. Review of the relations involved

### A) Zwanzig's expression

Several models have been developed to understand the complex dynamics, e.g. the random trap model[16], the random barrier model[17,18], random walk with barriers, continuous time random walk[8], etc. Theoretical analyses are mostly restricted to asymptotic long-time limits, when the particle motion should become diffusive. In an important treatment of the problem, Zwanzig[9] considered a general rough potential $U(x)$ with a smooth background $U_0(x)$ on which a perturbation $U_1(x)$ is superimposed, so that $U(x) = U_0(x) + U_1(x)$. He showed that the effective diffusion coefficient ($D_{eff}$) on the rough potential can be expressed as

$$D_{eff} = \frac{D_0}{\langle e^{\beta U_1} \rangle_{sp} \langle e^{-\beta U_1} \rangle_{sp}}, \qquad (4)$$

where $D_0$ is the bare diffusion coefficient on the smooth potential and $\langle \, \rangle_{sp}$ denotes the spatial average used to smooth the perturbation. He noted that this expression should also be valid in a random potential, where the amplitude of roughness has a Gaussian distribution,

$$P(U_1) = \frac{1}{\delta\sqrt{2\pi}} \left[ \exp\left(-\frac{U_1^2}{2\delta^2}\right) \right], \tag{5}$$

in which $\delta$ is the root-mean-squared roughness, $\delta^2 = \langle U_1^2 \rangle$. Hence, on such Gaussian random potential, Zwanzig's effective diffusion coefficient ($D_Z$) predicts strong dependence on temperature and roughness,

$$D_Z = D_0 \exp(-\beta^2 \delta^2). \tag{6}$$

Despite the novelty of the work and simplicity of the expression, the derivation of the above invokes, in the simplification of the double integral that arises in the evaluation of the mean first passage time (MFPT), a questionable local averaging of the random energy surface. Zwanzig himself was aware of the possible limitation of his approximate approach, and termed his final result as conjectural.

There are actually multiple unanswered issues in this problem. First and foremost, the existence of diffusion itself could be doubtful at large ruggedness. Imagine that the particle encounters a situation where it is stuck in a deep minimum (negative energy) with maxima (barriers, positive energy) on its two sides. Such a configuration becomes increasingly probable as ruggedness (that is, $\delta$) increases, and can give rise to long trapping and hence sub-diffusive growth of the mean square displacements. In fact, such three-site correlations get partially ignored in the coarse-graining mentioned above. We find that exact evaluation of the MFPT (by performing the multiple integrals involved) provides result that deviates from the one obtained by Zwanzig by coarse-graining of the potential. Second, the mean first passage time approach to estimate the diffusion constant (where $\tau_{MFPT}$ is compared with that of an effective flat potential, thereby implicitly invoking the relation $D_{eff} = L^2/2\tau_{MFPT}$ where $\tau_{MFPT}$ is the MFPT between an initial and final position separated by a distance $L$) might not work. Third, and a related issue (already discussed), is the question of ergodicity. The long trapping in the deep minima results in a "broken ergodicity" on such random potential energy surface, which have strong resemblance with the glass transition scenario. This is where the relationship between diffusion and entropy can have a role to play. Fourth, one needs to find an alternative coarse-graining / renormalization approach to describe the "broken ergodicity".

**B) Rosenfeld Scaling**

This well-known entropy-diffusion scaling relationship was first proposed by Rosenfeld in 1977 on the basis of earlier simulation results for the transport coefficients of a wide variety of one-component systems including those containing hard spheres, soft spheres, or plasma[6,7]. Using macroscopic reduction parameters for the length as $\rho^{-1/3}$ and the thermal velocity as $(k_B T/m)^{1/2}$, Rosenfeld demonstrated that, in dimensionless units, the self-diffusivity $D_0$ of a bulk fluid is well correlated with the excess entropy in terms of an exponential relation

$$D_R \equiv D_0 \frac{\rho^{1/3}}{(k_B T/m)^{1/2}} \approx a \exp(b S_{ex}) \ , \tag{7}$$

where $S_{ex} = \dfrac{S - S_{id}}{Nk_B}$ is the reduced (dimensionless) excess entropy per molecule, $a$ and $b$ are the constants which depend on the system, but $b$ shows weak variation. As discussed often, the validity of Rosenfeld's relation was established essentially by empirical means.

The purpose of this note is to present two derivations of Rosenfeld scaling. The first one, discussed briefly earlier, establishes a relation between entropy and dispersion of the random energy, and connects to Zwanzig's expression. In the process we note that Zwanzig's expression neglects, through a coarse-graining, any spatial correlation in energy between any two locations in space. The second derivation uses the expression for the mean first passage time (MFPT) and uses some results from the first derivation to obtain Rosenfeld's relation.

**III. Rosenfeld expression as a consequence of a lack of spatial correlations in Zwanzig model**

A d-dimensional random energy landscape with Gaussian distribution and without any spatial correlation allows an easy and exact derivation of the partition function, which leads us to the excess entropy. In the absence of any spatial correlation, the derivation is the same as a one dimensional case and the only parameter that enters the derivation is the number of lattice sites. Here we demonstrate a correlation between Zwanzig's expression for the effective diffusion coefficient and Rosenfeld scaling.

We start with the partition function ($Q$) for a single particle (as our particles are non-interacting) in the random energy surface characterized statistically by a Gaussian distribution only. In this case, we can obtain exact expressions of the Hemholtz free energy ($F$) and the entropy ($S$) per particle, as follows.

$$Q = \sum_{i,\{U_{1i}\}} P(U_{1i}) \exp\left(-\frac{U_{1i}}{k_B T}\right) = N_L \exp\left(\frac{\delta^2}{2k_B^2 T^2}\right) \quad . \tag{8}$$

where $N_L$ is the number of lattice sites. We have considered just one particle. In the above, we have replaced the dual sum over lattice sites and energy levels by noting the equivalence of all the sites and carrying out the average over the distribution of energy given by Eq. (5). Now, the following steps are self-explanatory

$$F = -k_B T \ln Q = -k_B T \ln N_L - \frac{\delta^2}{2k_B T} \tag{9}$$

$$S = -\left(\frac{dF}{dT}\right) = k_B \ln N_L - \frac{\delta^2}{2k_B T^2} \quad . \tag{10}$$

Hence the excess (dimensionless) entropy $S_{ex}$, for a single particle becomes,

$$S_{ex} = \frac{S - S_{id}}{k_B} = -\frac{\delta^2}{2k_B^2 T^2} \quad . \tag{11}$$

Random energy landscape with Gaussian distribution and the corresponding form of the entropy were observed in simulation of a model glass former[19]. If we now combine Zwanzig's expression [Eq. (6)] for diffusion in one dimensional rough potential with Eq. (11), we obtain the Rosenfeld's expression relating diffusion with entropy

$$D_R = a \exp(-b S_{ex}) = D_0 \exp\left(-\frac{\delta^2}{k_B^2 T^2}\right).$$  (12)

The Rosenfeld scaling parameters are given as $a = D_0, b = 2$. While Zwanzig's expression is strictly valid only for one dimensional diffusion, our derivation for the entropy of a rough landscape is valid for any dimension, of course only in the absence of any spatial correlation.

Note also that the following aspects. Rosenfeld's expression has been well-tested in three dimensions and seems to have a general validity[6-8,20]. It remains to be seen if the $\exp\left(-b\frac{\delta^2}{k_B^2 T^2}\right)$ form remains valid in higher dimension, with the parameter $b$ dependent on dimension $d$.

In the following we present a statistical mechanical derivation of Rosenfeld's scaling.

## IV. A statistical dynamical derivation of Rosenfeld's scaling

We first present an exact derivation but valid only in one dimension. Subsequently, in sub-section IV(B) we present a derivation based on effective medium approximation to the transition rates that is valid for arbitrary dimension.

### A) 1 dimension

We start with the following expression of the effective diffusion coefficient obtained from the mean first passage time (MFPT) distribution[21,22]

$$\frac{D_{eff}}{D_0} = \frac{L^2/2}{\left\langle \int_0^L dx \exp[U(x)/(k_B T)] \int_0^x dy [-U(y)/(k_B T)] \right\rangle}.$$  (13)

where boundaries are placed at the origin (x=0) and at x=L. In the absence of spatial correlation, the denominator of Eq. (13) can be expressed as $\langle \int_0^L dx \exp[U(x)/(k_B T)] \rangle \langle \int_0^x dy [-U(y)/(k_B T)] \rangle$. When roughness has a Gaussian distribution, spatial integration and ensemble average can be exchanged[23], $\int_0^L dx \langle \exp[U(x)/(k_B T)] \rangle \int_0^x dy \langle [-U(y)/(k_B T)] \rangle$. In the $L \to \infty$ limit, we can introduce a position dependent Hemholtz free energy $F(x)$ as

$$F(x) = -k_B T \ln \langle \int_0^x dy \exp[-U(y)/(k_B T)] \rangle . \tag{14}$$

The above definition of partition function and free energy become meaningful when (i) there is no bias in the potential U(y) and (ii) x is sufficiently large. We now use the above free energy to rewrite the denominator in the expression Eq. (13) of the effective diffusion coefficient as

$$\frac{D_{\text{eff}}}{D_0} = \frac{L^2/2}{\int_0^L dx \langle \exp[U(x)/(k_B T)] \rangle \exp[-F(x)/(k_B T)]} . \tag{15}$$

As discussed above, when the energy surface is Gaussian, the free energy of the ensemble of particles on the random energy surface of length $x$ is obtained as

$$-F(x) = -k_B T \ln(x) - \frac{1}{2} \frac{\delta^2}{k_B T} , \tag{16}$$

where $\delta^2 = \langle \delta U^2 \rangle$ and $\delta U = U - \langle U \rangle$. The corresponding entropy is obtained as,

$$S(x) = -\frac{dF(x)}{dT} = k_B \ln(x) - \frac{1}{2} \frac{\delta^2}{k_B T^2} . \tag{17}$$

By introducing the reduced excess entropy defined by

$$S_{\text{ex}}(x) = \frac{S(x) - S_{\text{id}}(x)}{k_B} , \tag{18}$$

where $S_{id}(x) = k_B \ln(x)$ is the entropy of the ensemble of particles on the flat energy surface, the free energy in Eq. (3) can be expressed as

$$F(x) = -k_B T \ln(x) + k_B T S_{ex}. \tag{19}$$

In terms of the free energy given by Eq. (19), Eq. (15) can be rewritten as,

$$\frac{D_{eff}}{D_0} = \frac{L^2/2}{\langle \int_0^L dx \exp[U(x)/(k_B T)] \rangle x \exp(-S_{ex})}. \tag{20}$$

When the ruggedness in the energy surface is Gaussian, we have,

$$\langle \exp[U(y)/(k_B T)] \rangle = \exp\left(\frac{1}{2}\frac{\delta^2}{(k_B T)^2}\right) = \exp(-S_{ex}).$$

Therefore, Eq. (20) can be transformed into

$$\frac{D_{eff}}{D_0} = \frac{L^2/2}{\int_0^L dx\, x} \exp(2S_{ex}) = \exp(2S_{ex}). \tag{21}$$

This is the Rosenfeld relation, again with $a=D_0$ and $b=2$, as observed in Sec.II.

## B) Derivation valid for arbitrary dimension d

Diffusion under a rugged landscape in any larger dimension might be different from that in 1 dimension[24]. For dimensions higher than one dimension, an exact calculation of the mean first passage time is not available, and therefore the approaches used above break down. Here, we derive the Rosenfeld scaling by employing the effective medium approximation. The effective medium theory is formulated using a lattice model, where transition rates between a pair of neighboring sites under rough potential are embedded into the effective medium. The effective transition rate is determined by requirering that the averaging the Green function over the different realizations of the rough potential embedded in the effective medium should reproduce that in the complete effective medium without embedded rough potential. By denoting the effective transition rate by $\Gamma_{eff}$, the effective diffusion constant can be obtained using Einstein's relation as $D = \Gamma_{eff} k_B T$. For the simplest case, a single transition rate is assumed to fluctuate by rough potential on two neighboring sites.[25] For the rate expression including the one we studied previously,[26] the activation energy for the fluctuating transition rate depends on the site energy difference alone. $\Gamma(\Delta U_i)$ denotes the transition from site i with a potential denoted by $U_i$ to a nearest neighbor site with a potential denoted by $U_j$ and $\Delta U_i = U_j - U_i$. The self-consistency condition in the effective medium theory requires symmetric transition rates.[22] The symmetric rates in view of the detailed balance can be given by[22]

$$\Gamma^{(sym)} = \rho_i^{(eq)} \Gamma(\Delta U_i) , \qquad (22)$$

where $\rho_i^{(eq)}$ is the equlibrium occupancy probability. The effective rate in the symmetric case can be defined using the original unsymmetrized rate $\Gamma_{eff}$ as $<\rho^{(eq)}> \Gamma_{eff}$. These are the same because $<\rho^{(eq)}> = 1$.

We consider the case that the ruggedness in the energy surface is Gaussian. As discussed above, the equilibrium occupation probability at site i denoted by $\rho_i^{(eq)}$ can be expressed as

$$\rho_i = \exp[-U_i/(k_B T)] / \langle \exp[-U_i/(k_B T)] \rangle = \exp[-U_i/(k_B T) + S_{ex}] , \qquad (23)$$

where the excess entropy is given by Eq. (11). In the effective medium theory, $\Gamma_{eff}$ is determined by the requirement that the effect of fluctuation on the site occupation probability for return to the starting site is zero on average. The self-consistency condition in d-dimension can be expressed as[22,25]

$$\left\langle \frac{\Gamma_{eff} - \Gamma^{(sym)}}{(d-1)\Gamma_{eff} + \Gamma^{(sym)}} \right\rangle = 0 , \qquad (24)$$

where *d* denotes the dimension. The self-consistent condition can be rewritten as

$$\frac{1}{d\Gamma_{eff}} = \left\langle \frac{1}{(d-1)\Gamma_{eff} + \Gamma^{(sym)}} \right\rangle. \tag{25}$$

We further rewrite the above equation as

$$\frac{1}{(d-1)}\left(1 - \left\langle \frac{\Gamma^{(sym)}}{(d-1)\Gamma_{eff} + \Gamma^{(sym)}} \right\rangle\right) = 0 \ . \tag{26}$$

We further rearrange the above equation to obtain,

$$\frac{1}{d} = \left\langle \frac{1}{1+(d-1)\Gamma_{eff}/\Gamma^{(sym)}} \right\rangle \ . \tag{27}$$

The above equation reduces to a trivial equality for d=1 but represents a very useful relation for $d>1$. This is because the factor $1/[1+(d-1)\Gamma_{eff}/\Gamma^{(sym)}]$ varies between zero and one in a way similar to Fermi-Dirac distribution function as we will show below.

In the absence of potential roughness, we have $\Delta U_i = 0$. In order to see the genuine effect of the potential roughness, we introduce a normalized transition rate defined by,

$$\Gamma_n(\Delta U_i) = \Gamma(\Delta U_i)/\Gamma(0) \ . \tag{28}$$

It is also convenient to define $G_{eff} = \Gamma_{eff}/\Gamma(0)$. $\Gamma^{(sym)}$ satisfies a relation given by

$$\Gamma^{(sym)}/\Gamma_{eff} = \Gamma_n(\Delta U_i)\exp[-U_i/(k_BT) - S_{ex}]/G_{eff} \ .$$

With these definitions, the consistency condition can be rewritten as

$$\frac{1}{d} = \left\langle \frac{1}{1+(d-1)\exp[(U_i - \mu(\Delta U_i))/(k_BT)]} \right\rangle, \tag{29}$$

where a quantity analogous to chemical potential is given by,

$$\mu(\Delta U_i) = -k_BT\ln\left[(d-1)G_{eff}/\Gamma_n(\Delta U_i)\right] + k_BTS_{ex} \ . \tag{30}$$

The quantity inside $\langle \cdots \rangle$ can be approximated as 1 when $U_i$ is smaller than $\mu(\Delta U_i)$ and decreases to zero as the value increases over that of $\mu(\Delta U_i)$. We also note

$$\langle\cdots\rangle=\int_{-\infty}^{\infty}d\Delta U_i\int_{-\infty}^{\infty}dU_i\frac{1}{2\pi\delta^2}\exp\left(-\frac{(U_i+\Delta U_i/2)^2}{\delta^2}-\frac{\Delta U_i^2}{4\delta^2}\right)\cdots. \tag{31}$$

The average with respect to $U_i$ is given by a Gaussian function whose maximum is at $-\Delta U_i/2$. We need different approximation to evaluate the integration with respect to $U_i$ depending on the value of the maximum given by $-\Delta U_i/2$ and $\mu(\Delta U_i)$. The condition $\mu(\Delta U_i)<-\Delta U_i/2$ can be expressed as

$$\ln\left[\frac{\Gamma_n(\Delta U_i)}{(d-1)G_{\text{eff}}}\right]<-\frac{\Delta U_i}{2k_BT}-S_{\text{ex}}. \tag{32}$$

It is more convenient to re-express the above relation as

$$\Gamma_n(\Delta U_i)<(d-1)G_{\text{eff}}\exp\left(-\frac{\Delta U_i}{2k_BT}-S_{\text{ex}}\right). \tag{33}$$

If the rate satisfies $\Gamma_n(\Delta U_i)\leq\exp\left(-\frac{\Delta U_i}{2k_BT}\right)$, the above equation holds for $d\geq 2$ (at least) when $\delta$ is small so that $G_{\text{eff}}\sim 1$. The rates studied previously by us (and simulations performed) satisfy this relation.[26] When $\mu(\Delta U_i)<-\Delta U_i/2$, we can employ the saddle point method and the double integration reduces to single integration

$$\frac{1}{d}=\int_{-\infty}^{\infty}d\Delta U_i\frac{1}{2\sqrt{\pi\delta^2}}\exp\left(-\frac{\Delta U_i^2}{4\delta^2}\right)\frac{1}{1+(d-1)G_{\text{eff}}\exp\left[-\Delta U_i/(2k_BT)-S_{\text{ex}}\right]/\Gamma_n(-\Delta U_i/2)}. \tag{34}$$

The self-consistency condition expressed by single integration reproduces the results obtained from the self-consistency condition expressed by double integration for the rates we have studied previously[26]. When $\delta$ is small we can again employ the saddle point method and obtain $1/d\sim 1/[1+(d-1)G_{\text{eff}}\exp(-S_{\text{ex}})]$, where we have used $\Gamma_n(0)=1$ using Eq. (28). We finally obtain a scaling relation,

$$\Gamma_{\text{eff}}/\Gamma(0)\approx\exp(S_{\text{ex}})=\exp\left(-\frac{\delta^2}{2(k_BT)^2}\right). \tag{35}$$

Here, we have assumed that $\Gamma_n(-\Delta U_i/2)$ is smoothly varying function around $\Delta U_i=0$. The rate we have studied previously[26] can be expressed as $\Gamma_n(\Delta U_i)=\exp\left[-(|\Delta U_i|+\Delta U_i)(2k_BT)\right]$ and $\exp\left[-\Delta U_i/(2k_BT)\right]/\Gamma_n(-\Delta U_i/2)$ has a cusp at $\Delta U_i=0$. As a result, the result of eq. (35) may not be fully accurate. Nevertheless, they roughly satisfy Rosenfeld's scaling relation

$$D_R = a\exp(-b\Delta S_{ex}) = D_0 \exp\left(-\frac{1}{2}\frac{\delta^2}{k_B^2 T^2}\right), \tag{36}$$

where the Rosenfeld scaling parameters are given as $a=D_0$, $b=1$. We have earlier obtained the *b=2* in one dimension, and here we now find $b=1$ for $d\geq 2$. The difference reflects the fact that Zwanzig's expression is strictly valid only for one dimensional diffusion. Since our derivation for the entropy of a rough energy landscape is valid for any dimension higher than 1, we obtain the diffusion-entropy scaling relation but the value of $b$ should be corrected to $b=2$ to describe diffusion under rough potential in $d\geq 2$.

We find no evidence of pathological behavior of diffusion in $d=1$ in comparison to that in $d\geq 2$, although we do find that the change of diffusion in going from $d=1$ to $d=2$ is larger than that from $d=2$ to $d=3$ or any other sequential changes. However, that is not unexpected. Thus, the issues raised by Newman and Stein[24] about pathology in $d=1$ remain unanswered.

**V. Breakdown of diffusion-entropy scaling due to correlations**

We have shown above that Rosenfeld's diffusion-entropy scaling is valid not only for one dimensional rugged potential energy surface, it also holds fairly rigorously for multi-dimensional potential energy surface (d>1). Experiments and computer simulations have shown that the said scaling holds well over a range of temperature and density in dense liquids. However, the scaling is found to breakdown in glassy liquids where Adam-Gibbs relation takes over. The breakdown may be attributed to the neglect of spatial and temporal correlations that the tagged molecule experiences when diffusing through the medium. In the present section we present an analysis where we use a previously published result to analyse the role of correlations in greater detail.[26]

We start with the one dimensional model and assume that the site energies on the lattice positions have Gaussian correlations in position

$$<U_i U_j> = \delta^2 \exp(-(i-j)^2/2\sigma^2) \quad , \tag{37}$$

where $\sigma$ gives a measure of correlation length. In the continuum limit, the above expression reduces to

$$<U(0)U(x)> = \delta^2 \exp(-(a^2 x^2/2\sigma^2)) \quad , \tag{38}$$

where $\sigma$ is the correlation length and $a$ is the lattice spacing. In this case, one can use the method of mean first passage time to obtain the following expression for the effective diffusion constant

$$D_{eff}(\delta,\sigma) = D_0 \exp(-\beta^2 \delta^2)/\left[1 + erf\left(\frac{\beta\delta}{2}\sqrt{1-\exp\left(-a^2/2\sigma^2\right)}\right)\right] \quad . \tag{39}$$

When the correlation length $\sigma$ goes to infinity, we get back Zwanzig's result and Rosenfeld diffusion-entropy scaling. In the opposite limit, we get a correction term discussed before. Since we know the entropy exactly in this system, the above expression for diffusion on a correlated landscape shows the inadequacy of the scaling relation.

**VI.   Conclusion**

Despite considerable efforts over many decades, we do not seem to have any widely accepted explanation of proposed relationships between diffusion and entropy. While Adam-Gibbs relation has been derived in terms of an impending entropy crisis at glass transition[4] or invoking nucleation of an entropic droplet model with radius dependent surface tension[5], a similar explanation for Rosenfeld has not been forthcoming. As already mentioned, there appears to exist a range where both the two expressions can correlate diffusion with respective entropy terms.

In this work, we present a derivation of Rosenfeld's diffusion-entropy scaling for a one dimensional and $d \geq 2$ rough potential. The derivation seems to be robust. In addition, the present study provides a clue to the crossover from Rosenfeld relation to that of Adam-Gibbs in terms of ergodicity breaking three site traps that become increasingly frequent at large ruggedness. Below we address this issue in more detail.

As already observed in our earlier work, Zwanzig's expression fails in the limit of large values of ruggedness parameter. The error can be traced back to the coarse graining that removes certain "pathological" configurations, like one deep minimum flanked by two high barriers, called three site trap (TST). *Such an arrangement, when coarse grained, reduces to a small barrier that misses the significant trapping effect of such a three site trap (TST).* It was shown in Ref. 26, that Zwanzig's expression can be modified to include both the effects of such coarse graining and also effects of spatial correlations.

**ACKNOWLEDGMENTS**

This work was supported by JSPS KAKENHI Grant Number 15K05406. BB thanks grants from DST (India) and Sir JC Bose Fellowship for partial support for this work. BB also thanks Professor Iwao Ohmine and Prof. Shinji Saito for hospitality at IMS, Okazaki, Japan, and IMS for partial support.

**REFERENCES**

1. R. Zwanzig J. Chem. Phys., **97**, 3587 (1992).

2. B. Bagchi, *"Molecular Relaxation in Liquids"*(Oxford, New York, 2012).

3. B. Bagchi, "*Water in Biological and Chemical Processes : From Strcture and Dynamics to Function*" (Cambridge, Cambridge, 2013).

4. M. Adam and J.H. Gibbs, J. Chem. Phys. **43**, 139 (1965).

5. X. Xia and P. G. Wolynes, Phys. Rev. Lett. **86**, 5526 (2001); Proc. Natl. Acad. Sci. U.S.A. **97**, 2990 (2000).

6. Y. Rosenfeld, Chem. Phys. Lett. **48**, 467 (1977).

7. Y. Rosenfeld, Phys. Rev. A **15**, 2545 (1977).

8. M. Agarwal, M. Singh, S. Sharma, M.P.Alam, and C. Chakravorty, J. Phys. Chem. B **114**, 6995 (2010).

9. R. Zwanzig, Proc. Natl. Acad. Sci. **85**, 2029 (1988).

10. S. Lifson and J.L. Jackson, J. Chem. Phys. **36**, 2410 (1962).

11. E.W. Montroll and G.H. Weiss, J. Math. Phys. **6**, 167 (1965).

12. A. Miller and E. Abraham, Phys. Rev. **120**, 745 (1960).

13. H. Frauenfelder, S.G. Silgar and P.G. Wolynes, **254**, 1598 (1991).

14. J. D. Bryngelson and P.G. Wolynes, J. Phys. Chem. **93**, 6902 (1989).

15. S.S. Plotkin, J. Wang and P.G. Wolynes, J. Chem. Phys. **106**, 2932 (1997).

16. Haus, J. W., Kehr, K. W. & Lyklema, J. W. Phys. Rev. B **25**, 2905 (1982).

17. J. Bernasconi, H. U. Beyeler, S. Strässler, S. Alexander, Phys. Rev. Lett. **42**, 819 (1979).

18. R. L Jack and P. Sollich, J. Stat. Mech. **2009**, 11011 (2009).

19. S. Sastry, Nature **409**, 164 (2001).

20. M. K. Nandi, A. Banerjee, S. Sengupta, S. Sastry, and S. M. Bhattacharyya, "Unraveling the success and failure of mode coupling theory from consideration of entropy" arXiv:1507.00203v1.

21. G. H. Weiss. Adv. Chem. Phys. **13,** 1 (1967).

22. K.W. Kehr, T. Wichmann, "Diffusion Coefficients of Single and Many Particles in Lattices with Different Forms of Disorder", Materials Science Forum, **223-224**, 151 (1996).

23. K. Seki and M. Tachiya, Phys. Rev. B **65**, 014305 (2001).

24. C. M. Newman and D. L. Stein, Ann. l'inst. Henri Poincarè (B) Probab. Stat. 31, 249 (1995), available online at http://www.numdam.org/item?id=AIHPB_1995__31_1_249_0.

25. S. Kirkpatrick, Rev. Mod. Phys. **45**, 574 (1973).

26. S Banerjee, R Biswas, K Seki, B Bagchi, J. Chem. Phys. **141,** 124105 (2015).